# Stimulated emission at 1.54 μm from Erbium/Oxygen-doped silicon-based light emitting diodes

JIN HONG[1,#], HUIMIN WEN[2,#], JIAJING HE[2], JINGQUAN LIU[2], YAPING DAN[2,*], JENS W. TOMM[3], FANGYU YUE[1,*], JUNHAO CHU[1,4], AND CHUNGANG DUAN[1,*]

[1]Key Laboratory of Polar Materials and Devices, Ministry of Education, East China Normal University, Shanghai 200241, China.

[2]National Key Laboratory of Science and Technology on Micro/Nano Fabrication Laboratory, Department of Micro/Nano Electronics, University of Michigan-Shanghai Jiao Tong University Joint Institute, Shanghai Jiao Tong University, Shanghai 200240, China.

[3]Max-Born-Institut für Nichtlineare Optik und Kurzzeitspektroskopie, Max-Born-Str. 2A, 12489 Berlin, Germany.

[4]National Laboratory for Infrared Physics, Shanghai Institute of Technical Physics, Chinese Academy of Sciences, 500 Yutian Road, Shanghai 200083, China.

[#]These authors contributed equally

[*]Corresponding authors: yaping.dan@sjtu.edu.cn (Yaping Dan), cgduan@clpm.ecnu.edu.cn (Chungang Duan), and fyyue@ee.ecnu.edu.cn (Fangyu Yue).

**Abstract**：Silicon-based light sources including light-emitting diodes (LEDs) and laser diodes (LDs) for information transmission are urgently needed for developing monolithic integrated silicon photonics. Silicon doped by ion implantation with erbium ions ($Er^{3+}$) is considered a promising approach, but suffers from an extremely low quantum efficiency. Here we report an electrically pumped superlinear emission at 1.54 μm from Er/O-doped silicon planar LEDs, which are produced by applying a new deep cooling process. Stimulated emission at room temperature is realized with a low threshold current of ~6 mA (~0.8 A/cm$^2$). Time-resolved photoluminescence and photocurrent results disclose the complex carrier transfer dynamics from the silicon to $Er^{3+}$ by relaxing electrons from the indirect conduction band of the silicon. This picture differs from the frequently-assumed energy transfer by electron-hole pair recombination of the silicon host. Moreover, the amplified emission from the LEDs is likely due to a *quasi-continuous* Er/O-related donor band created by the deep cooling technique. This work paves a way for fabricating superluminescent diodes or efficient LDs at communication wavelengths based on rare-earth doped silicon.

## 1. INTRODUCTION

Silicon/Si-based light sources including lasers at telecommunication wavelengths are the bottleneck for the heterogeneous integration of photonics with complementary metal-oxide-semiconductor circuits [1-6]. Ion implantation of Erbium/Er (often with Oxygen/O) into Si is believed to be one of the most promising approaches to create Si-based light emitting devices (LEDs) at 1.54 μm [7-15]. However, the reported quantum efficiencies are extremely low (≈ 0.01%) at room temperature, mainly due to strong non-radiative recombination caused by the comparably large Er-related precipitates formed during the cooling process in the *standard* rapid thermal annealing (RTA) [3,15-17]. Recently, the efficiency has been substantially improved by introducing a deep cooling (DC) technique [11,18] which can effectively mitigate the Er precipitation created during the RTA process. LEDs with a perpendicular emission structure based on the obtained material achieved a record external quantum efficiency of ~0.8% at room temperature [11].

Here, by optimizing the implantation of Er and O, the DC procedure, and the Si-based LED structure with a planar emission geometry, a near-unity quantum yield (or ~100% slope efficiency) of the photoluminescence (PL) in Er-doped Si is reached. Moreover, a super-linearly growing electroluminescence (EL) is obtained with a slope efficiency beyond 2 even at room temperature. The low threshold of ~6 mA (~0.8 A/cm$^2$) [19,20], the super-linear EL integral intensity [21-23], the narrowing full width at half maximum (FWHM) [24,25], and the Gaussian-like spatial emission distribution [26,27], as well as the fast radiation recombination lifetime [28-31], confirm the presence of stimulated emission involving transitions related to the Er$^{3+}$ [32-34]. Time-resolved PL (TR-PL) and photocurrent (PC) measurements reveal the relaxation dynamics of the non-equilibrium carriers from the Si host to the Er$^{3+}$ precipitates. The hot non-equilibrium carriers in Si first cross the intra-valley barrier (e.g., Γ or L point in $k$-space if the laser energy is high enough) with a time constant of ~110 ps to the bottom of the indirect conduction band (CB) (or the $\Delta_{1c}$ point) [35, 36]. The excess carriers further decay from the CB bottom to the evenly-distributed Er/O-related complexes that act as a quasi-continuous donor band with a decay time of ~30 ps. Within this donor band, the carriers resonantly excite (i.e., transfer the energy of carriers by non-radiative recombination to) the $Er^{3+}$-4$f$ electrons to emit at ~1.54 μm ($^4I_{13/2} \rightarrow {}^4I_{15/2}$).

## 2. MATERIALS AND METHODS

**A. Sample fabrication**

Float zone (FZ) intrinsic Si (100) wafers (Resistivity: ≥ 10 kΩ·cm; Thickness: 500 ± 20 μm; Suzhou Resemi Semiconductor Co., Ltd, China). Er and O ions were implanted with an injection energy and dose of 200 keV and 4 × $10^{15}$ $cm^{-2}$, and 32 keV and $10^{16}$ $cm^{-2}$, respectively, at the Institute of Semiconductors, Chinese Academy of Sciences, China. After that, the Er/O-implanted Si samples were cleaned with ethanol and deionized water, and then immersed in a piranha solution (sulfuric acid: 30% hydrogen peroxide=3:1) for 30 min at 90°C, followed by drying with a high purity nitrogen (99.99%) stream. To form a planar pn junction, we further implanted boron/B and phosphorus/P dopants into these Si samples (B: 30 keV and 2.2 × $10^{14}$ $cm^{-2}$; P: 80 keV and 2.2 × $10^{14}$ $cm^{-2}$). They have a similar peak concentration located ~80 nm below the surface, which matches the peak depth of the Er in Si. A 200-nm-thick $SiO_2$ films was then deposited on the Er/O-implanted Si samples by reactive magnetron sputtering (Delton multi-target magnetic control sputtering system, AEMD, SJTU). A deep cooling process was performed to activate the Er/O, B, and P dopants at the same time via an upgraded dilatometer (DIL 805A, TA Instruments) [11, 18], where the samples were annealed at 950°C for 5 min by means of copper coil-based electromagnetic heating and followed by a flush of high purity He (99.999%) gas cooled in liquid $N_2$ (77 K). Its detailed description can be referred to our previous works in Ref. 11.

A pair of co-axial electrodes was prepared by UV photolithography (MDA-400M, MIDAS) and metal film deposition (Nexdep, Angstrom Engineering Inc.). The internal electrode is in contact with *p*-type boron doping region and the external electrode in contact with the *n*-type P region. All the microfabrication processes were performed with home-built devices at the Center for Advanced Electronic Materials and Devices. After the Al metal wire bonding (7476D, West Bond), the devices were integrated on a PCB board. The *I-V* curves were taken using a digital sourcemeter (Keithley 2400) controlled by a LabVIEW script.

**B. Optical characterizations**

A Fourier transform infrared (FTIR) spectrometer (Vertex 80v, Bruker) is employed to measure transmission, reflectance, steady-state photoluminescence/PL and electroluminescence/EL, and photocurrent/PC spectra. The focused 405-nm emission from a continuous wave/cw LD (MLL-III-405, CNI, Changchun, China) with a maximum excitation power of ~160 mW was employed as the excitation source. The effective excitation power on the sample surface was calibrated by referring to the 405 nm-transmission and reflectance of the samples at different temperatures. Different optical filters including notch filters, neutral density filters, and long-pass filters were utilized for avoiding the influence of the excitation source during the excitation power dependent PL measurements. For EL measurements, cw and pulsed (with a ns-μs pulse duration and a repetition rate of ~$10^4$ Hz) currents were injected into the structures. For PC measurements, a low-noise current preamplifier (SRS SR570) was employed for recording the defect-related PC signal below the bandgap of Si. Time-resolved PL (TR-PL) was carried out in order to determine the non-equilibrium carrier lifetime related to $Er^{3+}$. This was implemented by mW-excitation with a 405 nm emitting diode laser being operated at 3 kHz as the excitation source, a fast InGaAs photodiode, and a GHz sampling oscilloscope (Agilent MSO9404A) for data accumulation and readout. Simultaneously, the PL decay curve was cross-checked by a luminescence spectrometer. TR-PL at an 80-MHz repetition rate was implemented by using a Tsunami Ti: sapphire laser (760 nm or 380 nm by using second harmonic, spot size ~100 μm) with 80-fs excitation pulses. The maximum energy density per pulse amounts to 57 and 3.4 μJ/cm$^2$ at 780 and 390 nm, respectively. Detection is made by a Hamamatsu C5680 streak camera with S1 photocathode operated in synchro-scan mode. The overall temporal resolution of the setup (1/e decay of an 80 fs-pulse) is better than 10 ps. In all setups the samples are mounted on the cold head of a Helium closed cycle cooler. This allows for temperature adjustment from 4 K to ambient.

Near Infrared EL imaging from the Er/O-Si LEDs was obtained at room temperature under optical microscope (BX53M, Olympus) equipped with a near-infrared camera (C12471-03, Hamamatsu). The recorded video is provided in the supplementary Video.

**3. RESULTS AND DISCUSSION**

Room temperature PL spectra of Er/O-doped Si (see Sample Information) at different excitation powers ($\lambda$ = 405 nm; 3.06 eV) are shown in Fig. 1a. Er$^{3+}$ emission at ~1.54 µm (~0.81 eV; marked with A) is observed and becomes more pronounced when the excitation power increases. Besides the main emission peak A, there are a side shoulder and a broad tail at the high energy side denoted as B (at ~0.82 eV) and C (even beyond the bandgap of Si), respectively (see inset). The shoulder B is ascribed to the $Er^{3+}$-related states [12] and the tail C is related to the Er/O-doping induced defect centers in the bandgap of Si. No shape change of the emission occurs when the excitation power is higher than ~5 mW. The integral intensity of the three specified emission bands A, B, and C shows a slope efficiency of near-unity ($S \approx 0.93$) within two orders of magnitude of excitation power (Fig. 1b in a log-log scale). It characterizes the quasi-free carrier recombination and the high internal quantum yield of $Er^{3+}$ without efficient (Auger or thermally-related) non-radiative recombination even at room temperature. At low temperature, the PL efficiency is enhanced as expected by suppressing the Shockley-Read-Hall recombination [11,15]. What's interesting is that the shoulder B and tail C are evident and can be strengthened in comparison with the main peak A as the excitation power increases; see PL spectra (4 K) at 10 mW and 100 mW in Fig. 1c. It is clear by comparing the curve at 4 K with 300 K that elevated temperatures have strong influence on the emission processes in B and C. Fig. 1d (normalized to peak A) demonstrates that the long tail C is substantially narrowed by shifting the edge-emission toward the long wavelength side upon increase of temperature; also see the two colored areas in Fig. 1c. The narrowing of the tail C unlikely originates from the temperature-dependent absorptivity of excitation light ($\lambda$ = 405 nm) as this effect has been taken into account in our measurements (see Fig. S1). Instead, it likely comes from the temperature-driven redistribution of excess carriers from the high-energy states in the Si CB to those within the Si bandgap (see more discussions later). This narrowing results in a relatively stronger emission for both shoulder B and tail C as shown in Fig. 1d.

To have a deeper insight into the above process, the integral intensity of the main peak A, the side shoulder B, and the tail C are normalized to their integral intensity at 4 K, respectively. The normalized integral intensities are shown in the Arrhenius plot in Fig. 1e. While the main peak A quenches monotonously with increase of temperature, emissions of B and C first increase and then decrease at a critical temperature point of ~180 K (~15.5 meV). Moreover, the total integral PL intensity also shows a monotonous decrease but with a relatively low temperature-quenching rate at 300 K (still ~50% of 4 K, i.e., with a decrease of 3 dB). This means that excitons transfer between the states, which are involved into the generation of the emission bands A, B, and C, without being affected by strong temperature-dependent non-radiative recombination. Based on the evolutions in Fig. 1e, an activation energy of ~14.9 meV can be extracted for the main peak A (rather close to the phonon energy of ~15.5 meV) [15], and ~40.0 meV for the total emission. Fig. 1f shows the PL decay of the 1.54 μm emission at 300 K and 77 K for an excitation power of ~20 W/cm$^2$. A comparable lifetime of ~14 μs can be obtained for both 300 K and 77 K by fitting their fast component curves. Notice that this value is at least one order of magnitude shorter than the ~ms lifetime of the Er$^{3+}$ in the literature [11,18,37], suggesting a different behavior of the emission compared to spontaneous emission.

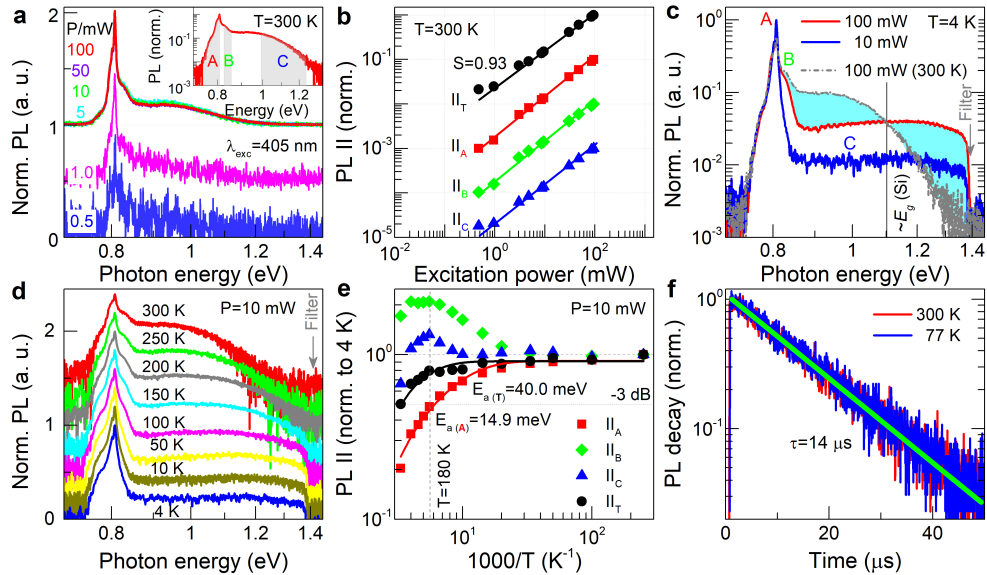

**Fig. 1 | PL spectra.** (**a**) Room-temperature PL spectra from the Er/O-doped Si at different excitation powers (*P*). PL curves are tentatively divided into three

parts, marked by A, B and C, as shown in the inset. (**b**) Excitation power dependence of PL integral intensity (II) at 300 K. (**c**) PL spectra at $P = 10$ mW (blue curve) and 100 mW (red curve) at 4 K. The room-temperature PL result at $P = 100$ mW (gray curve) is also given for comparison. (**d**) PL spectra at different temperatures ($P = 10$ mW). (**e**) Arrhenius plot of normalized PL integral intensity vs. inversed temperature. (**f**) PL decay curves at 77 K and 300 K together with the fitting results.

A lateral LED structure based on the DC-processed Er/O-doped Si structure is fabricated with a planar emission geometry (in contrast to our previous perpendicular structure) [11] (Fig. 2a). We separately implanted boron/B and phosphorus/P into the Er/O-Si samples both with a peak concentration of $10^{19}$ cm$^{-3}$, forming a coaxial pn-junction diode on the Si surface. The DC process was then applied to activate the Er/O, B, and P dopants (see Sample Information). A rectifying current vs. voltage (*I-V*) curve was observed. Fig. 2b shows the EL spectra of the LED at different injection currents (at 300 K). Notice that the EL is similar in spectral shape to the PL from the Er-activated Si except that the tail C is weaker and cut off at the Si bandgap; see the curves at the bottom of Fig. 2c. With the injection current increase, the emission signal at ~0.81 eV becomes clearer with the feature of the Er$^{3+}$ emission at ~1.54 μm, and simultaneously narrowed first and then broadened. The EL cutoff at the Si bandgap (see the blue vertical line) is due to the fact that electrons can only be electrically pumped to the CB bottom. In PL, electrons can be excited to high energy states in the CB, resulting in a long emission tail well beyond the Si bandgap.

To clarify further the energetic structure in the LEDs, we recorded the short-circuit PC spectra at 300 K under light illumination from 0.6 eV to 1.3 eV in photon energy, as shown in Fig. 2c. The third-order derivative of the curve sets the cutoff at ~1.1 eV which corresponds to the Si bandgap $E_g$. It is necessary to point out that around 0.1 eV below the Si bandgap, the PC reaches a plateau after an exponential decline by 2 orders of magnitude, which extends to ~0.6 eV before it drops below the background noise. This PC plateau roughly lies in the same spectral range of the broad EL band, and should come from the Er-related defect (including $Er^{3+}$-4f degenerated) states [38].Taking into account the PC and EL bands from 0.7 eV to 1.1 eV, this implies that a quasi-continuous band from 0.7 eV above the valence band (VB) to the bottom of (even above) CB is formed.

Fig. 2d and e shows the main peak positions and FWHMs as a function of the injection current at 300 K and 7 K. It can be observed that the main emission peak at 1.54 μm is almost independent of the injection current, whereas the FWHM is reduced from 120 meV to 20 meV at 300 K (from 110 meV to 10 meV at 7 K) until the current approaches to ~10 mA. After a saturation region, the FWHM starts to broaden again when the injection current is higher than ~30 mA. Moreover, in Fig. 2f and g, as the injection current ramps up, the peak intensity and integral intensity distinctively show a sub-linear increase followed by super-linear ramp with a slope of $S \approx 2.7$ at 300 K (~3.6 at 7 K) and $S \approx 1.5$ at 300 K (almost the same at 7 K), respectively. A well-pronounced threshold is observed at ~6 mA or 0.8 A/cm$^2$ (refer to the integral intensity in Fig. 2g). These features along with the FWHM narrowing suggest that amplified spontaneous [21,23,24] (or stimulated [33,34]) emission occurs in our Si LEDs. The emission cross-section can be thus estimated by the following equation [30,39],

$$\sigma_{em} = \sqrt{\ln\left(\frac{2}{\pi}\right)} (\lambda^4 / 4\pi c n^2 \tau w) \qquad (1)$$

where $c$ is the light speed, $n$ is the refractive index (3.47), $\tau$ is lifetime (14 μs), $w$ is the FWHM (~0.10 μm) and $\lambda$ is the emission wavelength (1.54 μm). Its maximum is approximately $\sigma_{em} = 7.0 \times 10^{-19}$ cm$^2$, which is even one order of magnitude larger than the values for the Er$^{3+}$ in nanocrystal-Si sensitized silica (8.0 × 10$^{-20}$ cm$^2$) [15, 22].

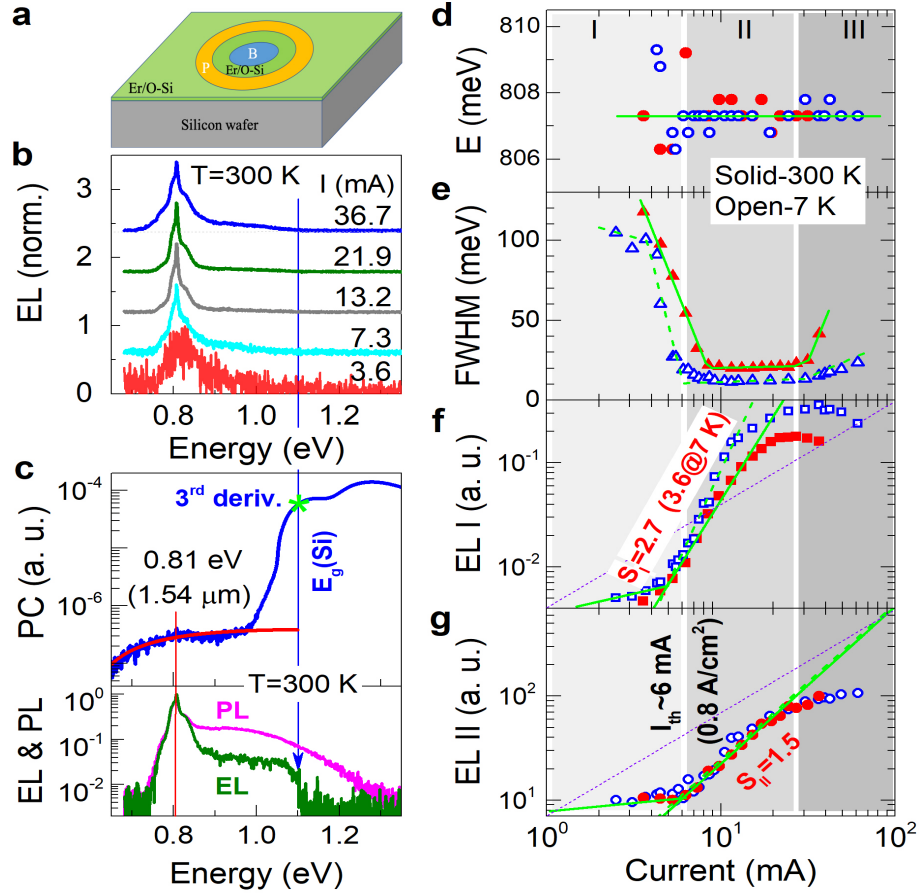

**Fig. 2 | EL spectra and analysis.** (**a**) Scheme of the LED structure. (**b**) Current-dependent EL spectra at 300 K. (**c**) PC spectrum at 300 K (upper panel). The star '*' marks the maximum of the third-order derivative and the blue line $E_g$ (Si) = 1.10 eV. The PC shoulder at ~0.81 eV is from the Er-related defects. The lower panel shows the EL and PL spectra of the LED at 300 K. (**d**)-(**g**) Current-dependence of the EL peak energy, FWHM, intensity and integrated intensity, respectively, at 300 K (red) and 7 K (blue). The green lines are guides to the eyes.

Fig. 3 shows the spatially-resolved emission intensity of the LED surface as taken by an InGaAs camera. At an injection current lower than the threshold (Fig. 3a), the emission is weak and uniformly distributed across the entire emitter surface (300 μm in diameter) except for the region close to the center electrode (see the 2-dimensional image at the bottom), which is heated up. When the injection current is higher than the threshold, the emission is strongly enhanced with a maximum in the LED surface center; see Fig. 3b. In this case, the spatial distribution of the emission follows a Gaussian distribution if the loss caused by the center electrode is not considered (the pixels are saturated in Fig. 3b). The supplementary Video.mp4 shows the evolution of the emission intensity profile as the LED is electrically pumped across the threshold current. Together with the above results including the behavior of the emission energy, FWHM, intensity, and integral intensity, this observation confirms the presence of electrically-driven stimulated emission, although it lacks a suitable cavity [32-34].

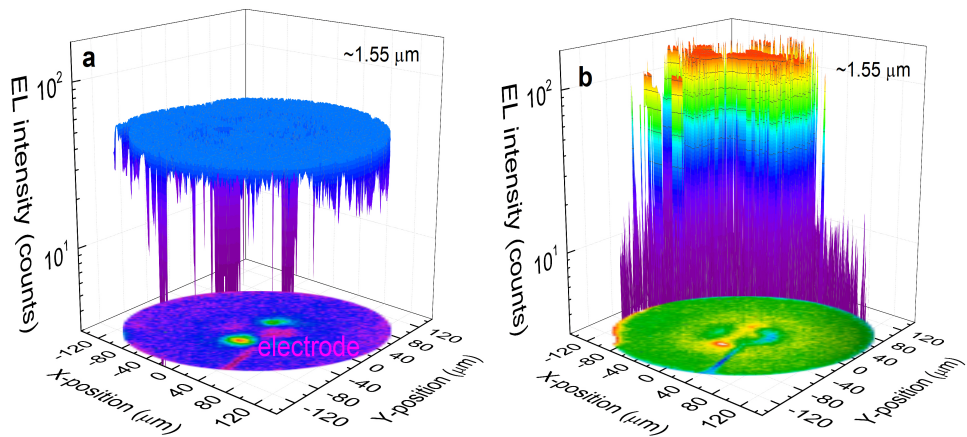

**Fig. 3 | Spatial-resolved intensity distribution at the emitting surface of the LED at 1.55 µm (a) below and (b) above the threshold current.** The emission imaging of the surface is also shown at the bottom for comparison.

We now address the electronic structure of the Er$^{3+}$-doped Si and the carrier relaxation and transfer dynamics. Fig. 4a shows the TR-PL results under a pulsed laser excitation ($\lambda$ = 760 nm, 80 MHz) at 5 K. The recorded PL signals are between 1.1 eV and 1.2 eV. On the time scale, the PL amplitudes remain nearly constant at a scale of ~12 ns, indicating that the emission has a longer lifetime. The time-integrated PL spectrum can be fitted with a Gaussian distribution with a peak centered at ~1.15 eV, as shown in Fig. 4b, which likely comes from the band-to-band emission in Si [40]. The corresponding decay profile shown in Fig. 4c suggests that the lifetime is much longer than ~12 ns, consistent with the picture presented in Fig. 4a. Notice that the streak camera used is not capable of detecting light near ~1.5 µm in wavelength, and that the time resolution is better than 10 ps; also see Fig. 4c.

When the frequency-doubled wavelength (380 nm) excites the sample, a similar but smaller emission band at ~1.15 eV was observed; see Fig. 4d. In the time domain, the emission has two striking features shown in Fig. 4e. First, unlike the long emission lifetime excited at the wavelength of 760 nm, the emission has a short decay time of ~30 ps. Second, the emission lags behind the excitation pulse by ~110 ps. These features are likely caused by the fact that high energy photons from the 380 nm laser can excite electrons from the VB to the L (or Γ') point of the CB, which the 760 nm laser excitation cannot. With assistance of phonons, hot electrons from the L point will first transit to the CB bottom ($\Delta_{1c}$ point) and then emit photons at ~1.15 eV via radiative recombination from band to band. The time constant of ~110 ps could be attributed to the transfer time of the carriers across the intra-valley barrier. The relatively long transit time results in a low concentration of excess carriers at the CB bottom because the barrier temporally holds the carriers in the L point. As a result, excess carriers do not generate strong spontaneous emission but rapidly (~30 ps) relax to the Er/O-related quasi-continuous levels below the indirect CB of Si [15,16]; see more discussion later. This behavior is illustrated by the blue transients in Fig. 4e. This fast relaxing process suggests a fundamental discrepancy with the existing model of energy transfer from electron-hole pair recombination directly to the excitation of the $Er^{3+}$-4f electrons for the 1.54 μm emission.

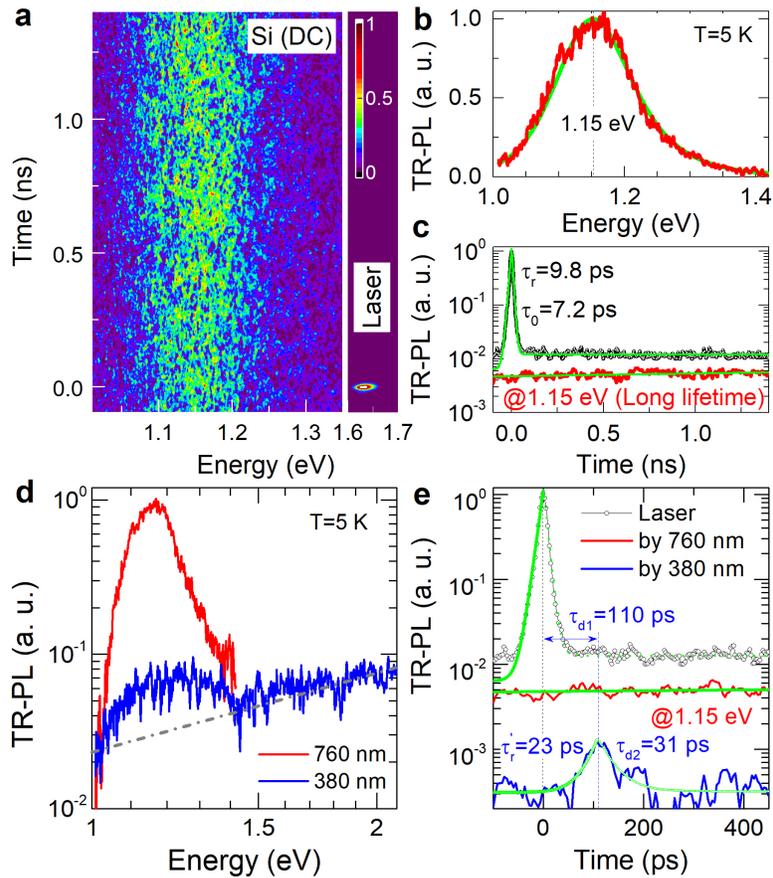

**Fig. 4 | TR-PL spectral results.** (**a**) TR-PL image from the Er/O-doped Si at 5 K (left) with 760-nm excitation laser. (**b**) Time-integrated spectrum and Gaussian-fit. (**c**) Decay curves of the sample and the laser. (**d**) Time-integrated spectra at excitation wavelengths of 760 nm and 380 nm. (**e**) Decay curves at different excitation wavelengths.

The analysis of all these findings results in a comprehensive picture, which is shown schematically in Fig. 5. High energy photons from the 380 nm laser first excite electrons from the valence to the L point of the CB. Due to the intervalley barriers, these carriers slowly transit to the bottom of the indirect CB ($\Delta_{1c}$) with a time constant of ~110 ps. Since the localized states in the quasi-continuous

Er/O-related donor band have larger momentum, excess carriers at the indirect CB bottom can be readily coupled and transferred to the Er/O-related donor band with a relatively short lifetime of ~30 ps. Finally, the carriers relax to the VB via radiative and nonradiative recombination in the quasi-continuous Er/O-related donor band. The radiative recombination broadens the emission band up to the Si bandgap and the nonradiative recombination mainly transfers energy to resonantly excite the $Er^{3+}$. Thus the quasi-continuous band not only facilitates the rapid emission decay of ~30 ps (by 380 nm laser excitation) shown in Fig. 4e, but also serves as efficient non-radiation recombination centers to resonantly excite the $Er^{3+}$ for the 1.54 µm emission. This could be one of the reasons why the 1.54 µm emission in Er/O-doped Si is usually not effectively excited by infrared photons, but by high-energy photons, e.g., ~400 nm, which then produce the amplified emission. Although the fast relaxation of carriers within the quasi-continuous donor band cannot be directly detected, the extremely long lifetime of the $Er^{3+}$ emission at 1.54 µm (~ms) facilitates the achievement of inversion, resulting in amplified spontaneous (or stimulated) emission. The main reason why it is complicated to obtain amplified emission by optical excitation (i.e., PL) is the high reflectance loss of visible light at the surface of the polished Si matrix. Although the temperature shows no influence on the reflectivity, the loss of surface reflectivity is high (e.g., ~40% for 405 nm) [41]; also see Fig. S1.

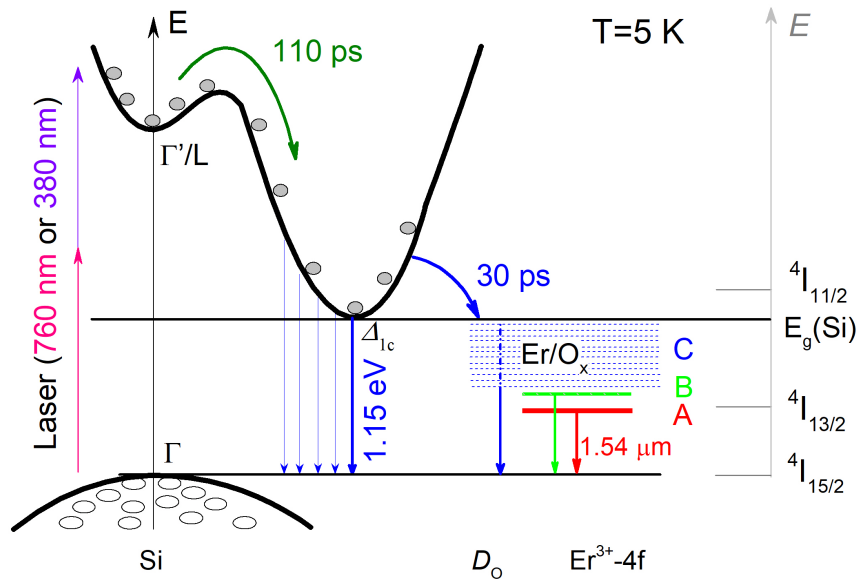

**Fig. 5 | Scheme of the carrier relaxation dynamics.** The hot-carriers in the upper states in Γ′ or L points excited by the 380 nm laser transfer to the indirect CB minimum ($\Delta_{1c}$) of Si with a time constant of ~110 ps. From here, a time constant of ~30 ps characterizes the transition of carriers to the distributed band created by the Er/O-related donor states.

What needs to be emphasized again is that the results achieved are largely due to the use of the DC process in sample preparation. The process offers advantages for the optically pumped $Er^{3+}$. These ions appear in high density in relatively uniformly sized, Er-related clusters with a diameter of ~1 nm, which is much smaller than the 5 nm diameter achieved with the standard RTA process [11]. Moreover, the spatial distribution of the small clusters is extremely even, which provides very useful conditions for the Er/O cluster to act as a broad and quasi-continuous donor band [42]. The decaying carriers from the Si host into the Er-related states excite the 4f-electron of $Er^{3+}$, which produces the emission at ~1.54 μm. Details on material characterizations by high resolution transmission electron microscopy and X-ray photoelectron spectroscopy have been described in Ref. 11, and the striking difference of the defect-related EL signal between RTA and DC is shown in Fig. S2.

## 4. CONCLUSION

In conclusion, we observed amplified spontaneous (or stimulated) emissions at room temperature with a low threshold of ~0.8 A/cm$^2$ from the Er/O-doped Si-based LEDs that were treated with a DC process. In comparison with the standard RTA process, the DC process can effectively suppress the precipitation of Er/O related nanocrystals and form more uniformly distributed Er-O-Si compounds. The impact of the DC process on Er/O formation can effectively reduce the density of nonradiative defects in the Si bandgap and facilitate the formation of a quasi-continuous Er/O-related donor band right below the Si CB. As a result, strong room temperature PL and amplified spontaneous (or stimulated) emissions were observed in EL spectra. In particular, the quasi-continuous Er/O-related donor band not only facilitates the rapid emission decay of excess carriers but also serves as efficient recombination centers to extend the emissions up to the Si bandgap, forming a broad tail in the spectrum in addition to the widely observed $Er^{3+}$ emission at 1.54 μm. This work may pave a way for fabricating superluminescent (or laser) diodes at communication wavelengths based on rare-earth doped silicon.

**Funding.**

National Science Foundation of China (61874043, 61790583, 21703140, 61874072); Aero-Science Fund (201824X001); Special-key project of the "Innovative Research Plan", Shanghai Municipality Bureau of Education (2019-01-07-00-02-E00075).

**Acknowledgment.**

The authors would like to thank Dr. Ren Zhu for the characterization of spatial-resolved intensity distribution at the emitting surface of the LED

**Disclosures.**

The authors declare no conflicts of interest.

**REFERENCES**

1. X. Chen, M. M. Milosevic, S. Stanković, S. Reynolds, T. D. Bucio, K. Li, D. J. Thomson, F. Gardes, and G. T. Reed, "The emergence of silicon photonics as a flexible technology platform," Proc. IEEE 106, 2101-2116 (2018).

2. D. Thomson, A. Zilkie, J. E. Bowers, T. Komljenovic, G. T. Reed, L. Vivien, D. Marris-Morini, E. Cassan, L. Virot, and J.-M. Fédéli, "Roadmap on silicon photonics," J. Opt. 18, 073003 (2016).

3. Z. Zhou, B. Yin, and J. Michel, "On-chip light sources for silicon photonics," Light Sci. Appl. 4, e358 (2015).

4. H. Subbaraman, X. Xu, A. Hosseini, X. Zhang, Y. Zhang, D. Kwong, and R. T. Chen, "Recent advances in silicon-based passive and active optical interconnects," Opt. Express 23, 2487-2511 (2015).

5. D. Liang and J. E. Bowers, "Recent progress in lasers on silicon," Nat. Photonics 4, 511-517 (2010).

6. D. A. Miller, "Device requirements for optical interconnects to silicon chips," Proc. IEEE 97, 1166-1185 (2009).

7. E. M. Fadaly, A. Dijkstra, J. R. Suckert, D. Ziss, M. A. van Tilburg, C. Mao, Y. Ren, V. T. van Lange, K. Korzun, and S. Kölling, "Direct-bandgap emission from hexagonal Ge and SiGe alloys," Nature 580, 205-209 (2020).

8. C. Roques-Carmes, S. E. Kooi, Y. Yang, A. Massuda, P. D. Keathley, A. Zaidi, Y. Yang, J. D. Joannopoulos, K. K. Berggren, and I. Kaminer, "Towards integrated tunable all-silicon free-electron light sources," Nat. Commun. 10, 1-8 (2019).

9. S. Chen, W. Li, J. Wu, Q. Jiang, M. Tang, S. Shutts, S. N. Elliott, A.

Sobiesierski, A. J. Seeds, and I. Ross, "Electrically pumped continuous-wave III–V quantum dot lasers on silicon," Nat. Photonics 10, 307 (2016).

10. Y. Takahashi, Y. Inui, M. Chihara, T. Asano, R. Terawaki, and S. Noda, "A micrometre-scale Raman silicon laser with a microwatt threshold," Nature 498, 470-474 (2013).

11. H. Wen, J. He, J. Hong, S. Jin, Z. Xu, H. Zhu, J. Liu, G. Sha, F. Yue, and Y. Dan, "Efficient Er/O-doped silicon light-emitting diodes at communication wavelength by deep cooling," Adv. Opt. Mater. 8, 2000720 (2020).

12. M. A. Hughes, H. Li, N. Theodoropoulou, and J. D. Carey, "Optically modulated magnetic resonance of erbium implanted silicon," Sci. Rep. 9, 1-10 (2019).

13. M. Celebrano, L. Ghirardini, M. Finazzi, G. Ferrari, Y. Chiba, A. Abdelghafar, M. Yano, T. Shinada, T. Tanii, and E. Prati, "Room temperature resonant photocurrent in an erbium low-doped silicon transistor at telecom wavelength," Nanomaterials 9, 416 (2019).

14. M. Lourenço, M. Milošević, A. Gorin, R. Gwilliam, and K. Homewood, "Super-enhancement of 1.54 μm emission from erbium co-doped with oxygen in silicon-on-insulator," Sci. Rep. 6, 1-6 (2016).

15. A. Kenyon, "Erbium in silicon," Semicond. Sci. Tech 20, R65 (2005).

16. N. Q. Vinh, N. N. Ha, and T. Gregorkiewicz, "Photonic properties of Er-doped crystalline silicon," Proc. IEEE 97, 1269-1283 (2009).

17. G. Franzò, F. Priolo, and S. Coffa, "Understanding and control of the erbium non-radiative de-excitation processes in silicon," J. Lumin. 80, 19-28 (1998).

18. Wen H, Hong J, Yue F & Dan Y. Optical characterization and optimization of near-infrared emission by erbium doping in silicon. (to be published).

19. H. Shen, D.-S. Li, and D.-R. Yang, "Research progress of silicon light

source," Acta Phys. Sin. 64, 204208 (2015).

20. S. Wu, S. Buckley, J. R. Schaibley, L. Feng, J. Yan, D. G. Mandrus, F. Hatami, W. Yao, J. Vučković, and A. Majumdar, "Monolayer semiconductor nanocavity lasers with ultralow thresholds," Nature 520, 69-72 (2015).

21. J. Hong, H. Wang, F. Yue, J. W. Tomm, D. Kruschke, C. Jing, S. Chen, Y. Chen, W. Hu, and J. Chu, "Emission kinetics from PbSe quantum dots in glass matrix at high excitation levels," Phys. Status Solidi-R 12, 1870312 (2018).

22. J. Ramírez, F. F. Lupi, Y. Berencén, A. Anopchenko, J. Colonna, O. Jambois, J. Fedeli, L. Pavesi, N. Prtljaga, and P. Rivallin, "Er-doped light emitting slot waveguides monolithically integrated in a silicon photonic chip," Nanotechnology 24, 115202 (2013).

23. T. Kobayashi, M. Djiango, and W. J. Blau, "Near-infrared electroluminescence and stimulated emission from semiconducting nonconjugated polymer thin films," J. Appl. Phys. 107, 023103 (2010).

24. V. Ho, Y. Wang, B. Ryan, L. Patrick, H. Jiang, J. Lin, and N. Vinh, "Observation of optical gain in Er-doped GaN epilayers," J. Lumin. 221, 117090 (2020).

25. N. T. Otterstrom, R. O. Behunin, E. A. Kittlaus, Z. Wang, and P. T. Rakich, "A silicon Brillouin laser," Science 360, 1113-1116 (2018).

26. H. Liu, Z. Li, W. Song, Y. Yu, F. Pang, and T. Wang, "$MoS_2$/graphene heterostructure incorporated passively mode-locked fiber laser: from anomalous to normal average dispersion," Opt. Mater. Express 10, 46-56 (2020).

27. S. Saito, T. Takahama, K. Tani, M. Takahashi, T. Mine, Y. Suwa, and D. Hisamoto, "Stimulated emission of near-infrared radiation in silicon fin light-emitting diode," Appl. Phys. Lett. 98, 261104 (2011).

28. V. Robbiano, G. M. Paternò, A. A. La Mattina, S. G. Motti, G. Lanzani, F. Scotognella, and G. Barillaro, "Room-temperature low-threshold lasing from monolithically integrated nanostructured porous silicon hybrid microcavities," ACS Nano 12, 4536-4544 (2018).

29. K. Luterová, D. Navarro, M. Cazzanelli, T. Ostatnický, J. Valenta, S. Cheylan, I. Pelant, and L. Pavesi, "Stimulated emission in the active planar optical waveguide made of silicon nanocrystals," Phys. Status Solidi C 2, 3429-3434 (2005).

30. A. Rapaport and M. Bass, "The role of stimulated emission in luminescence decay," J. Lumin. 97, 180-189 (2002).

31. M. Bresler, O. Gusev, E. Terukov, I. Yassievich, B. Zakharchenya, V. Emel'yanov, B. Kamenev, P. Kashkarov, E. Konstantinova, and V. Y. Timoshenko, "Stimulated emission in erbium-doped silicon structures under optical pumping," Mater. Sci. Eng. B 81, 52-55 (2001).

32. L. A. Coldren, S. W. Corzine, and M. L. Mashanovitch, Diode lasers and photonic integrated circuits (John Wiley & Sons, 2012), Vol. 218.

33. V. Ho, T. Al Tahtamouni, H. Jiang, J. Lin, J. Zavada, and N. Vinh, "Room-temperature lasing action in GaN quantum wells in the infrared 1.5 μm region," ACS Photonics 5, 1303-1309 (2018).

34. M. Huda and S. Ali, "A study on stimulated emission from erbium in silicon," Mater. Sci. Eng. B 105, 146-149 (2003).

35. M. Schultze, K. Ramasesha, C. Pemmaraju, S. Sato, D. Whitmore, A. Gandman, J. S. Prell, L. Borja, D. Prendergast, and K. Yabana, "Attosecond band-gap dynamics in silicon," Science 346, 1348-1352 (2014).

36. J. Noffsinger, E. Kioupakis, C. G. Van de Walle, S. G. Louie, and M. L.

Cohen, "Phonon-assisted optical absorption in silicon from first principles," Phys. Rev. Lett. 108, 167402 (2012).

37. X. Wang, B. Wang, L. Wang, R. Guo, H. Isshiki, T. Kimura, and Z. Zhou, "Extraordinary infrared photoluminescence efficiency of $Er_{0.1}Yb_{1.9}SiO_5$ films on $SiO_2$/Si substrates," Appl. Phys. Lett. 98, 071903 (2011).

38. B. De Geyter, A. J. Houtepen, S. Carrillo, P. Geiregat, Y. Gao, S. ten Cate, J. M. Schins, D. Van Thourhout, C. Delerue, and L. D. Siebbeles, "Broadband and picosecond intraband absorption in lead-based colloidal quantum dots," ACS Nano 6, 6067-6074 (2012).

39. W. J. Miniscalco and R. S. Quimby, "General procedure for the analysis of $Er^{3+}$ cross sections," Opt. Lett. 16, 258-260 (1991).

40. W. L. Ng, M. Lourenco, R. Gwilliam, S. Ledain, G. Shao, and K. Homewood, "An efficient room-temperature silicon-based light-emitting diode," Nature 410, 192-194 (2001).

41. I. Costa, D. Pera, and J. A. Silva, "Improving light capture on crystalline silicon wafers," Mater. Lett., 127825 (2020).

42. M. Bürkle, M. Lozac'h, C. McDonald, M. Macias‐Montero, B. Alessi, D. Mariotti, and V. Švrček, "Tuning the bandgap character of quantum-confined Si-Sn alloyed nanocrystals," Adv. Funct. Mater. 30, 1907210 (2020).